\documentclass[doublecol]{epl2} 

\usepackage{amsmath,amssymb}
\usepackage{color}

\def\ad{{a^\dagger}}

\title{Quantum simulations of light-matter interactions in arbitrary coupling regimes}
\shorttitle{Quantum simulations of light-matter interactions} 

\author{L. Lamata\inst{1}}
\shortauthor{L. Lamata}

\institute{                    
  \inst{1} Departamento de F\'isica At\'omica, Molecular y Nuclear, Universidad de Sevilla, 41080 Sevilla, Spain
}

\abstract{
Light-matter interactions are an established field that is experiencing a renaissance in recent years due to the introduction of exotic coupling regimes. These include the ultrastrong and deep strong coupling regimes, where the coupling constant is smaller and of the order of the frequency of the light mode, or larger than this frequency, respectively. In the past few years, quantum simulations of light-matter interactions in all possible coupling regimes have been proposed and experimentally realized, in quantum platforms such as trapped ions, superconducting circuits, cold atoms, and quantum photonics. We review this fledgling field, illustrating the benefits and challenges of the quantum simulations of light-matter interactions with quantum technologies.}

\begin{document}

\maketitle

\section{Introduction } Light-matter interactions have existed as a growing field inside physics for more than a century~\cite{Loudon}. In addition to the exciting theoretical and experimental advances in this area that have found applications in quantum technologies~\cite{Nielsen}, a novel avenue has emerged that brings fresh perspectives to the field, namely, light-matter interactions in the ultrastrong and deep-strong coupling regimes~\cite{Forn-Diaz,Kockum}. In these regimes, the light-matter coupling is of the order of the electromagnetic mode frequency and smaller than it, in the first case, and larger than the mode frequency in the second case. These are regimes hard to attain in atomic systems, but they have been achieved in some cases in superconducting circuits as well as in polaritons. For detailed reviews on the field, see Refs.~\cite{Forn-Diaz,Kockum}.

Quantum simulators are controllable quantum systems capable of reproducing the features of other quantum systems, such as dynamics and static properties, e.g., the ground state energy~\cite{QSimNori}. They have been proposed and experimentally implemented in a variety of simulated topics, such as high-energy physics, quantum chemistry, and condensed-matter physics, and quantum platforms, as for example trapped ions, superconducting circuits, cold atoms, and quantum photonics.

In the past few years, the quantum simulation of light-matter interactions in all possible coupling regimes has been proposed and implemented, either in trapped ions~\cite{Pedernales,Puebla_16,Aedo,Kim,TwoPhotonRabi,RabiQPT,Cheng}, superconducting circuits~\cite{Ballester,Mezzacapo,Lamata,Langford}, cold atoms~\cite{Felicetti}, as well as photonics~\cite{Crespi}.

In this Perspective, we start with a revision of the quantum Rabi model from quantum optics, describing the coupling between a two-level quantum system and a quantized bosonic mode. Later on, we give an overview of the research on quantum simulations of light-matter interactions, focusing on three paradigmatic examples: the quantum simulation of the quantum Rabi model in a trapped ion~\cite{Pedernales}, a proposal to implement the Dicke model with a multi-ion system~\cite{Aedo}, as well as the quantum Rabi and Dicke models in superconducting circuits~\cite{Mezzacapo}. Moreover,  in the last section we will briefly describe further results in other quantum platforms such as photonics~\cite{Crespi} and cold atoms~\cite{Felicetti}.

\section{Light-matter interaction: the quantum Rabi model}

The quantum Rabi model (QRM) is the basic model of light-matter interactions~\cite{Rabi}. It consists of a two-level spin coupled to an electromagnetic bosonic mode. When it is extended to several spins coupled to the same mode, the result is called the Dicke model~\cite{Dicke}. The Hamiltonian for the QRM takes the form, in the Schr\"odinger picture  ($\hbar=1$),
\begin{equation}
H_{\rm QRM}= \frac{\omega_0}{2}\sigma_z + \omega a^\dag a + g\sigma_x (a^\dag+a)\label{QRM1},
\end{equation}
where $\omega_0$ is the spin frequency, $\omega$ is the electromagnetic mode frequency, $g$ is the light-matter coupling, $a^\dag$, $a$ are creation and annihilation operators of the electromagnetic bosonic mode, and $\sigma_{x,z}$ are Pauli matrices.

Typically, in atomic systems the coupling $g$ is several orders of magnitude smaller than the mode frequency $\omega$ and the spin frequency $\omega_0$, such that an approximation can be made. This is called the rotating-wave approximation (RWA), and consists of neglecting the counterrotating terms, depending on $\omega_0+\omega$, in the interaction Hamiltonian in an interaction picture with respect to $(\omega_0/2)\sigma_z+\omega a^\dag a$,
\begin{equation}
H_{\rm I}^{\rm i.p.}=g(\sigma^+ a \exp[i(\omega_0-\omega)t]+\sigma^+ a^\dag \exp[i(\omega_0+\omega)t] + H.c.).
\end{equation}
The result is the Jaynes-Cummings (JC) model~\cite{Jaynes}, which can be easily solved as it enjoys a $U(1)$ symmetry, and in the Schr\"odinger picture takes the form,
\begin{equation}
H_{\rm JCM}= \frac{\omega_0}{2}\sigma_z + \omega a^\dag a + g(\sigma_+ a+\sigma_- a^\dag)\label{JCM1}.
\end{equation}

When extended to several spins coupled to the same mode, the Jaynes-Cummings model transforms into the Tavis-Cummings model~\cite{Tavis}, which is nothing but the Dicke model under small coupling $g$ as compared to $\omega$ and $\omega_0$, after applying the rotating-wave approximation.

Two interesting results in this area are the analytical solution to the quantum Rabi model in all coupling regimes, obtained in recent years~\cite{Braak}, as well as the exploration of coupling regimes beyond the rotating-wave approximation~\cite{Casanova}, namely, the ultrastrong coupling regime (USC, $0.1 \omega  \lesssim g  \lesssim \omega$), and the deep-strong coupling regime (DSC, $g \gtrsim  \omega$). Further recent results in the literature involve the exploration of quantum phase transitions in the quantum Rabi model~\cite{RabiQPT}.

\section{Quantum Rabi model with trapped ions}

Here we will review the proposal in Ref.~\cite{Pedernales} for implementing the QRM in all parameter regimes with a single trapped ion. This proposal has been experimentally carried out in Ref.~\cite{Kim}, successfully reproducing the dynamics and ground state properties.

Individual ionized atoms confined employing Paul traps can have their motional quantum states cooled down to their ground state employing laser cooling techniques~\cite{InnsbruckReview}. Moreover, two internal electronic levels of a trapped ion can be employed as a quantum spin. Via red and blue sideband interactions, one may couple the internal spin and the bosonic motional mode, with the Hamiltonian 
 \begin{eqnarray}
 \nonumber
 H & = & \frac{\nu_0}{2} \sigma_z + \nu a^\dag a + \Omega(\sigma_+ +  \sigma_-) \Big({\rm exp}\{i[\eta(a + a^\dag)\\
 &-& \omega_lt + \phi_l]\} +  {\rm exp}\{-i[\eta(a + a^\dag) -\omega_l t + \phi_l]\} \Big).
\end{eqnarray}
Here, $\sigma_+$ and $\sigma_-$ are the raising and lowering Pauli operators, $a^\dag$ and $a$ are the creation and annihilation operators of the motional mode,  $\nu$ is the motional mode frequency, $\nu_0$ is the spin transition frequency, $\eta$ is the Lamb-Dicke parameter, $\Omega$ is the Rabi interaction strength, while $\omega_l$ and $\phi_l$ are, respectively, the frequency and phase of the laser driving. Under a bichromatic driving, moving into an interaction picture with respect to the free-energy Hamiltonian, ${H_0 = \frac{\nu_0}{2}\sigma_z} + \nu a^\dag a $ while applying an optical RWA, the Hamiltonian results~\cite{NISTreview03}
\begin{eqnarray}
H^{\rm I} & = & \!\! \sum_{n=r,b}\frac{\Omega_n}{2}\left[e^{i\eta[a(t)+a^\dag(t)]}e^{i(\nu_0-\omega_n)t}\sigma_++\textrm{H.c.}\right],
\label{trapped_ion_hamil}
\end{eqnarray}
where  $a(t)=ae^{-i\nu t}$ and $a^\dag (t)=a^\dag e^{i\nu t}$. 
The frequencies of the red-sideband (r) and blue-sideband (b) drivings, with detunings $\delta_r$ and $\delta_b$, read,
\begin{eqnarray}
\omega_r = \nu_0+\nu+\delta_r , \,\,\, \omega_b = \nu_0-\nu+\delta_b . \nonumber
\end{eqnarray}
Neglecting fast oscillating terms in Eq.~(\ref{trapped_ion_hamil}), while considering the Lamb-Dicke regime, i.e., $\eta\sqrt{\langle a^\dag a\rangle}\ll1$ we select terms that oscillate with minimum frequency, assuming $\delta_n,\Omega_n\ll\nu$ for $n = r , b$. With these approximations we obtain the simplified Hamiltonian
\begin{equation}
\bar{H}^{\rm I} = \frac{i\eta\Omega}{2} \sigma_+ \left( a e^{-i\delta_rt} + a^\dag e^{-i\delta_bt} \right) + \textrm{H.c.},
\end{equation}
where we assume equal interaction strengths for both sidebands, $\Omega = \Omega_r = \Omega_b$.
By moving into an additional interaction picture with respect to $H_0 = \frac{1}{4}(\delta_b+\delta_r)\sigma_z + \frac{1}{2}(\delta_b-\delta_r) a^\dag a $, we eliminate the time dependence in $\bar{H}^{\rm I} $ and the quantum Rabi Hamiltonian is obtained,
\begin{equation}
\bar{H}_{\rm QRM} = \frac{\omega_0^R}{2}\sigma_z + \omega^R a^\dag a + ig (\sigma^+-\sigma^-)(a+a^\dag) ,
\label{effective_hamiltonian}
\end{equation} 
with the simulated spin and bosonic mode frequencies
\begin{eqnarray}
\omega_0^R=-\frac{1}{2}(\delta_r+\delta_b), \,\, \omega^R = \frac{1}{2}(\delta_r-\delta_b), \,\, g=\frac{\eta\Omega}{2} 
\end{eqnarray}
provided by the sum and difference of both detunings, respectively. Given the fact that these parameters can be tuned at will in a trapped-ion experiment, this allows one for the study of the QRM in all possible coupling regimes via the appropriate choice of the ratio $ g / \omega^R$. It is notable that all carried out interaction-picture mappings are of the form $\alpha a^\dag a +\beta\sigma_z$. This formula commutes with the standard observables in the QRM, $\{ \sigma_z, | n \rangle \langle n |, a^{\dagger} a \}$, such that the experimental measurements will not be affected by the transformations.

\subsection{Access to exotic regimes---} Here we will describe the regimes which are difficult to achieve with atomic systems without a quantum simulation, namely, the ultrastrong and deep-strong coupling ones. For a complete description of the quantum simulation with a trapped ion of all possible coupling scenarios, we refer to the original Ref.~\cite{Pedernales}.

The USC regime is defined in the interval $0.1 \lesssim g/\omega^R \lesssim 1$, with perturbative and nonperturbative limits. In this range of parameters, the RWA does not hold even with the spin on resonance with the mode. In this situation, the dynamics has to be described in terms of the complete quantum Rabi Hamiltonian. In the case where $g/\omega^R \gtrsim 1$, we reach the DSC region, for which the evolution can be explained in terms of phonon number wave packets bouncing back and forth along well established parity chains~\cite{Casanova}.

The access to all possible regimes is constrained by the maximum detunings permitted for the drivings, which are provided by the relation $\delta_{r,b} \ll \nu$. This will allow that high-order sidebands are not populated. In Ref.~\cite{Pedernales}, a numerical simulation of the full Hamiltonian in Eq.~(\ref{trapped_ion_hamil}) was carried out with typical ion-trap parameters: $\nu=2\pi \times 3 {\rm MHz}$, $\Omega=2\pi \times 68 {\rm kHz}$ and $\eta=0.06$~\cite{Gerritsma10}, with laser detunings given by $\delta_b=- 2\pi \times 102 {\rm kHz}$ and $\delta_r=0$. This would correspond to a simulation of the JC regime with $g/\omega^{\rm R}=0.01$, which, interestingly, is the most demanding regime for the quantum simulator, as it has the largest detunings, even though it is analytically solvable. The numerical simulations showed that second-order sidebands were not populated and that the state dynamics followed the analytical JC solution with a fidelity above $99\%$ for many Rabi oscillations. Values of $\eta=0.06$ also allow for up to few tens of phonons, enabling the Lamb-Dicke regime and allowing for high-fidelity quantum simulation of the QRM in all possible coupling scenarios.

With respect to decoherence times, the relevant timescale of the quantum simulation corresponds to $t_{\rm char}= \frac{2\pi}{g}$. In the proposed quantum simulator, $g=\frac{\eta \Omega}{2}$, such that $t_{\rm char}=\frac{4 \pi}{\eta \Omega}$. For realistic values of $\eta=0.06-0.25$ and of $\Omega/2\pi=0-500$~kHz, the dynamics timescale of the considered system is of the order of milliseconds, being well below decoherence times of ionic electronic levels and motional degrees of freedom~\cite{InnsbruckReview}.

\subsection{Ground state preparation---} The ground state $|G\rangle$  of the QRM in the JC regime ($g\ll\omega^R$) corresponds to the state $|g,0\rangle$, namely, the spin ground state, $| g \rangle$, as well as the bosonic mode vacuum, $| 0 \rangle$. It is well known that $|g,0\rangle$ does not provide the ground state of the interacting system for larger light-matter coupling regimes, for which the contribution of the counter-rotating terms is sizable~\cite{hwang2010}. As described in Ref.~\cite{Pedernales} and experimentally demonstrated in Ref.~\cite{Kim}, the ground state of the USC and DSC Hamiltonians is highly nontrivial~\cite{Braak}, mainly because it contains spin and mode excited states, i.e., $\langle G|a^\dag a|G\rangle>0$.

Therefore, initializing the spin-bosonic mode system in its real ground state is a quite involved state-engineering exercise in most parameter situations, except for the JC limit. In Ref.~\cite{Pedernales} it was proposed to generate the ground state of the USC/DSC coupling regimes of the QRM via an adiabatic evolution. When the QRM system is prepared in the JC region, achieved, e.g., with detunings $\delta_r=0$ and $|\delta_b |\gg g$, it represents a JC Hamiltonian with its ground state corresponding to $|G\rangle=|g,0\rangle$. We point out that the $g/\omega^{\rm R}$ ratio can be slowly increased, adiabatically mapping the system in the configuration space to higher coupling regions~\cite{Kyaw2014}. This may be achieved in two ways, namely, increasing $g$ by a larger intensity of the driving, or decreasing $\omega^{\rm R}$ by a smaller detuning $|\delta_b|$. The adiabatic theorem~\cite{messiah} guarantees that, for a slow enough process, Landau-Zener transitions to excited states will not happen, the system continuing in its ground state.  

\section{Quantum Simulation of the Dicke Model in Trapped Ions}

We review now how to simulate the Dicke model in a linear ion trap, following the proposal in Ref.~\cite{Aedo}. The Dicke model \cite{Dicke}, being the natural extension of the quantum Rabi model \cite{Rabi, Braak}, consists of $N$ two-dimensional spins coupled to a single electromagnetic field bosonic mode. The interaction term between the spins and the bosonic mode can be expressed in terms of a Tavis-Cummings plus an anti-Tavis-Cummings terms, producing the following Hamiltonian,
\begin{align}\label{Dicke_Hamiltonian}
H_{\text{D}}=&\omega \ad a+\sum_{m=1}^N\frac{\omega_m^q}{2}\sigma_m^z+\nonumber\\
+&\overbrace{\sum_{m=1}^N g_m (a\sigma_m^++\text{H.c.})}^{\text{Tavis-Cummings}}+\overbrace{\sum_{m=1}^N g_m (\ad \sigma_m^++\text{H.c.})}^{\text{anti-Tavis-Cummings}}.
\end{align}

Moving into an interaction picture with respect to the uncoupled Hamiltonian, $\omega \ad a+\sum_{m=1}^N\omega_m^q\sigma_m^z/2$ one obtains,
\begin{align}
H_{\text{D}}^{\text{I}}=&\sum_{m=1}^N g_m\left(a \sigma_m^+ e^{i(\omega_m^q-\omega)t}+\text{H.c.}\right)+\nonumber\\
+&\sum_{m=1}^Ng_m\left(\ad \sigma_m^+ e^{i(\omega_m^q+\omega)t}+\text{H.c.}\right).
\end{align}
Considering all ion frequencies equal and the bosonic field mode to be coupled with the same interaction strength to all qubits, namely, $\omega_m^q=\omega^q$ and $g_m=g$ ($\forall m\in[1,N]$), the Hamiltonian results
\begin{align}\label{Dicke_interaction_picture}
H_{\text{D}}^{\text{I}}=&g\left(a \Sigma^+ e^{i(\omega^q-\omega)t}+\text{H.c.}\right)+\nonumber\\
+&g\left(\ad \Sigma^+ e^{i(\omega^q+\omega)t}+\text{H.c.}\right).
\end{align}

Following the protocol developed in Ref.~\cite{Pedernales} for the QRM, as described in previous section, namely, the single-spin Dicke model, the crucial feature to implement a controllable quantum model with an analog quantum simulator is to recognize the similarities between that model and the trapped-ion device. A similar situation will happen in the multiqubit case, as described in Ref.~\cite{Aedo}. The Hamiltonian for the Dicke model in Eq. (\ref{Dicke_interaction_picture}) corresponds to the sum of red-sideband and blue-sideband Hamiltonians of a multi-ion system, straightforwardly generalized from the formalism in the previous section, i.e., $H_{\text{D}}^{\text{I}}=H^{\text{r}}+H^{\text{b}}$, as long as the following choice is made
\begin{equation}\label{parameters_Dicke}
g=\frac{\Omega\eta}{2}\ , 
\phi^\text{r}=\phi^\text{b}=-\frac{\pi}{2}\ , 
\delta^\text{r}=\omega-\omega^q\ , 
\delta^\text{b}=-(\omega+\omega^q)\ ,
\end{equation}
where we assume the Rabi frequencies for red and blue sideband to be equal, $\Omega^\text{r}=\Omega^\text{b}=\Omega$. Accordingly, the simulated frequencies of the Dicke model can be expressed in terms of the motional mode frequency $\nu$, the frequency of the two-level spins $\omega^0$, as well as the laser frequencies $\omega^\text{r}$ and $\omega^\text{b}$,
\begin{align}
\omega&=\frac{\delta^\text{r}-\delta^\text{b}}{2}=\nu+\frac{\omega^\text{r}-\omega^\text{b}}{2}\nonumber\\
\omega^q&=-\frac{\delta^\text{r}+\delta^\text{b}}{2}=\omega^0-\frac{\omega^\text{r}+\omega^\text{b}}{2},
\end{align}
as described in Ref.\cite{Aedo}. For the measurement and ground state initialization, one may proceed in a similar way as for the single-ion case~\cite{Pedernales,Aedo}.

\section{Quantum Rabi and Dicke models with superconducting circuits}

An analog quantum simulator of the quantum Rabi model with superconducting circuits was introduced in Ref.~\cite{Ballester}. However, this analysis is tricky to be carried out experimentally due to the large drivings required. We will now review the digital-analog proposal for the quantum simulation of the quantum Rabi and Dicke models with superconducting circuits introduced in Ref.~\cite{Mezzacapo}. The experimental implementation of this result was carried out in Ref.~\cite{Langford}. An extension to generalized Dicke models was put forward in Ref.~\cite{Lamata}. The formalism employed in this proposal is the corresponding to a digital-analog quantum simulator: a combination of analog blocks (in this case, JC interactions) with digital steps (in this case, single-spin rotations). The paradigm of digital-analog quantum simulation and computation was reviewed in Ref.~\cite{DAQS}.

In this case, the blue-sideband interaction will not be as easily available as in trapped ions, such that we will follow a digital-analog approach~\cite{DAQS}, considering only red sideband (JC) drivings and single-spin rotations. We begin by assuming a standard circuit QED setup with a charge-like qubit, e.g. a transmon qubit~\cite{Koch07}, which will play the role of our spin, being coupled to a mw resonator. This scenario follows the Hamiltonian~\cite{Blais04} 
\begin{equation}
H=\omega_r a^{\dagger}a +\frac{\omega_q}{2}\sigma_z +g(a^{\dagger}\sigma_-+a\sigma_+),\label{QubitResHam}
\end{equation}
where $\omega_q$ and $\omega_r$ are the qubit (i.e., spin), and resonator transition frequencies, $g$ is the spin-resonator interaction strength, $a^{\dagger}$ and $a$ are the creation and annihilation operators for the resonator mode, respectively, and $\sigma_{\pm}$ are the raising and lowering spin operators.

The proposal in Ref.~\cite{Mezzacapo} is based on realizing that the evolution of the quantum Rabi Hamiltonian 
\begin{equation}
H_R=\omega^R_r a^{\dagger}a +\frac{\omega^R_q}{2}\sigma_z +g^R\sigma_x(a^{\dagger}+a)\label{RabiHam}
\end{equation}
can be simulated with a  superconducting circuit device obeying a JC interaction, such as in Eq.~(\ref{QubitResHam}), by means of a digital decomposition. 

The quantum Rabi Hamiltonian in Eq.~(\ref{RabiHam}) can be expanded in two terms, $H_R=H_1+H_2$, where
\begin{eqnarray}
&&H_1=\frac{\omega^R_r}{2} a^{\dagger}a +\frac{\omega^1_q}{2}\sigma_z +g(a^{\dagger}\sigma_-+a\sigma_+) , \nonumber \\
&&H_2=\frac{\omega^R_r}{2} a^{\dagger}a -\frac{\omega^2_q}{2}\sigma_z +g(a^{\dagger}\sigma_++a\sigma_-) ,
\label{Ham12} 
\end{eqnarray}
being the spin transition frequency defined for the two steps such that $\omega_q^1-\omega_q^2=\omega^R_q$. These two evolutions can be carried out in a superconducting circuit system with fast control of the spin (i.e., qubit) transition frequency. If we start from the spin-resonator Hamiltonian in Eq.~(\ref{QubitResHam}), we can employ a picture rotating at frequency $\tilde{\omega}$, with which the resulting interaction Hamiltonian reads 
\begin{equation}
\tilde{H}=\tilde{\Delta}_ra^{\dagger}a+\tilde{\Delta}_q\sigma_z+g(a^{\dagger}\sigma_-+a\sigma_+),\label{IntHam}
\end{equation}  
where $\tilde{\Delta}_r=(\omega_r-\tilde{\omega})$ and $\tilde{\Delta}_q=\left(\omega_q-\tilde{\omega}\right)/2$. 
Accordingly, Eq.~(\ref{IntHam}) coincides with $H_1$, after a suitable redefinition of the parameters.
The anti-JC, counterrotating term $H_2$ can be carried out via a local spin rotation to $\tilde{H}$ as well as another detuning for the spin transition frequency,
\begin{equation}
e^{-i \pi\sigma_x/2}\tilde{H}e^{i \pi\sigma_x/2}=\tilde{\Delta}_ra^{\dagger}a-\tilde{\Delta}_q\sigma_z+g(a^{\dagger}\sigma_++a\sigma_-).\label{RotHam}
\end{equation}

If we choose different spin-resonator detuning for the two parts, $\tilde{\Delta}^1_q$ in the first case and $\tilde{\Delta}^2_q$ in the second one, we will be able to quantum simulate the quantum Rabi Hamiltonian, Eq.~(\ref{RabiHam}), by means of a digital-analog decomposition~\cite{DAQS}, by successive iterations of the interleaved interactions.  By increasing the number of digital steps, one may improve the fidelity via reducing the digital error.

The Rabi parameters of the simulated model will be mapped to the physical parameters of the quantum simulator according to the expressions, $\omega_r^R=2\tilde{\Delta}_r$, $\omega_q^R=\tilde{\Delta}_q^1-\tilde{\Delta}_q^2$, $g^R=g$. 

To assess how realistic would the quantum simulation be, in Ref.~\cite{Mezzacapo} numerical simulations were carried out with realistic superconducting circuit parameters~\cite{Koch07}.
A frequency of $\omega_r/2\pi=7.5$~GHz, and a transmon-resonator coupling of $g/2\pi=100$~MHz were considered. The spin frequency $\omega_q$ and the rotating frame frequency $\tilde{\omega}$ were suitably varied to achieve different parameter scenarios.

In Ref.~\cite{Langford}, an experimental demonstration of this proposal was carried out. The number of digital (i.e., Lie-Trotter-Suzuki)~\cite{DAQS} steps was significantly high, up to 90, resulting in a large fidelity and an accurate reproduction of the model.

\section{Additional results in the literature} Without the goal of being exhaustive, here we describe some other articles which appeared in the literature on quantum simulations of light-matter interactions with quantum technologies. One of the first experimental realizations of a quantum simulation of the quantum Rabi model was carried out in a photonics platform~\cite{Crespi}. In this experiment, an analog quantum simulator of the quantum Rabi model was performed, by means of light transport in femtosecond-laser-written waveguide superlattices. This approach allowed one for an experimentally accessible framework to explore the features of light-matter interaction in the ultrastrong and deep-strong coupling regimes. 

In Ref.~\cite{Felicetti}, a method was developed to carry out a quantum simulation of the quantum Rabi model with cold atoms in tailored optical lattices. In this work, an impementation of the two-level spin was proposed in terms of the population of Bloch bands in the first Brillouin zone. The effective spin would couple to a quantum bosonic mode provided by an optical dipole trap.

\section{Conclusions and future perspectives} In this article we have reviewed the field of quantum simulations of light-matter interactions with quantum technologies. We have described three prototypical results in the literature: the quantum simulation of the quantum Rabi model with a single trapped ion, a proposal for the quantum implementation of the Dicke model with trapped ions, and the quantum simulation of the quantum Rabi and Dicke models with superconducting circuits.

Quantum technologies may allow one to explore exotic regimes of light-matter interactions which are inaccesible to atomic systems, such as the ultrastrong and deep-strong coupling regimes. Even though these ranges of parameters may be simulated ab initio in quantum platforms such as superconducting circuits and condensed matter polaritons, it can be extremely useful to study these regimes in controllable quantum platforms such as quantum simulators. In these devices, the parameters can be tuned at will, and the cases to be analyzed can be much larger than with a fixed sample built in a nanofabrication lab facility. Moreover, even though the quantum simulations carried out in this area so far are straightforwardly simulatable with classical computers, future extensions of these pioneering works may involve scalability to many spins, as well as nonsymmetric situations including inhomogeneous couplings, bias terms, as well as pulsed dynamics. These cases may enable one to go beyond what is efficiently computable classically, and outperform the fastest classical supercomputers for exploring these nonstandard regions.

Finally, future explorations of extreme light-matter interactions with quantum simulators may enable the design of novel materials and devices, which could have an impact in science, technology, and industry.

\acknowledgments
We acknowledge the funding by PGC2018-095113-B-I00, PID2019-104002GB-C21, and PID2019-104002GB-C22 (MCIU/AEI/FEDER, UE).


\begin{thebibliography}{0}



\bibitem{Loudon}
  \Name{Loudon R.}
  \Book{The Quantum Theory of Light}
  \Publ{Oxford University Press, Oxford, UK}
  \Year{2000}.

\bibitem{Nielsen}
  \Name{Nielsen M. A. \and Chuang I. L.}
  \Book{Quantum Computation and Quantum Information}
  \Publ{Cambridge University Press, Cambridge, UK}
  \Year{2000}.
  
\bibitem{Forn-Diaz}
  \Name{Forn-D\'iaz P., Lamata L., Rico E., Kono J., \and Solano E.}
  \REVIEW{Rev. Mod. Phys.}{91}{2019}{025005}.

\bibitem{Kockum}
  \Name{Frisk Kockum A., Miranowicz A., De Liberato S., Savasta S., \and Nori F.}
  \REVIEW{Nat. Rev. Phys.}{1}{2019}{19}.

\bibitem{QSimNori}
  \Name{Georgescu I. M., Ashhab S., \and Nori F.}
  \REVIEW{Rev. Mod. Phys.}{86}{2014}{153}.

\bibitem{Pedernales}
  \Name{Pedernales J. S., Lizuain I., Felicetti S., Romero G., Lamata L., \and Solano E.}
  \REVIEW{Sci. Rep.}{5}{2015}{15472}.

\bibitem{Puebla_16}
  \Name{Puebla, R., Casanova, J., \and Plenio, M. B.}
  \REVIEW{New J. Phys.}{18}{2016}{113039}.

\bibitem{Aedo}
  \Name{Aedo I. \and Lamata L.}
  \REVIEW{ Phys. Rev. A}{97}{2018}{042317}.

\bibitem{Kim}
  \Name{Lv D., An S., Liu Z., Zhang J.-N., Pedernales J. S., Lamata L., Solano E., \and Kim K.}
  \REVIEW{Phys. Rev. X}{8}{2018}{021027}.

\bibitem{TwoPhotonRabi}
  \Name{Felicetti S., Pedernales, J. S., Egusquiza, I. L., Romero, G., Lamata, L., Braak, D., and  Solano, E.}
  \REVIEW{Phys. Rev. A}{92}{2015}{033817}.

\bibitem{RabiQPT}
  \Name{Puebla, R., Hwang, M.-J., Casanova, J., and Plenio, M. B.}
  \REVIEW{Phys. Rev. Lett.}{118}{2017}{073001}.

\bibitem{Cheng}
  \Name{Cheng X.-H., Arrazola I., Pedernales J. S., Lamata L., Chen X., \and Solano E.}
  \REVIEW{Phys. Rev. A}{97}{2018}{023624}.

\bibitem{Ballester}
  \Name{Ballester, D., Romero, G., Garc\'ia-Ripoll, J. J., Deppe, F., \and Solano, E.}
  \REVIEW{Phys. Rev. X}{2}{2012}{021007}.

\bibitem{Mezzacapo}
  \Name{Mezzacapo A., Las Heras U., Pedernales J. S., DiCarlo L., Solano E., and Lamata L.}
  \REVIEW{Sci. Rep.}{4}{2014}{7482}.

\bibitem{Lamata}
  \Name{Lamata L.}
  \REVIEW{Sci. Rep.}{7}{2017}{43768}.

\bibitem{Langford}
  \Name{Langford, N. K., Sagastizabal, R., Kounalakis, M., Dickel, C., Bruno, A., Luthi, F., Thoen, D. J., Endo A., \and DiCarlo, L.}
  \REVIEW{Nat. Commun.}{8}{2017}{1715}.

\bibitem{Felicetti}
  \Name{Felicetti S., Rico E., Sabin C., Ockenfels T., Koch J., Leder M., Grossert C., Weitz M., and Solano E.}
  \REVIEW{Phys. Rev. A}{95}{2017}{013827}.

\bibitem{Crespi}
  \Name{Crespi A., Longhi S., \and Osellame R.}
  \REVIEW{Phys. Rev. Lett.}{108}{2012}{163601}.
  
  
  \bibitem{Rabi}
  \Name{Rabi, I. I.}
  \REVIEW{Phys. Rev.}{49}{1936}{324}.

 
\bibitem{Dicke}
  \Name{Dicke, R. H.}
  \REVIEW{Phys. Rev.}{93}{1954}{99}.

\bibitem{Jaynes}
  \Name{Jaynes, E. T. \and Cummings, F. W.}
  \REVIEW{Proc. IEEE}{51}{1963}{89}.

\bibitem{Tavis}
  \Name{Tavis, M. \and Cummings, F. W.}
  \REVIEW{Phys. Rev.}{170}{1968}{379}.
  
  \bibitem{Braak}
  \Name{Braak, D.}
  \REVIEW{Phys. Rev. Lett.}{107}{2011}{100401}.
 
  \bibitem{Casanova}
  \Name{Casanova, J., Romero, G., Lizuain, I., Garc\'ia-Ripoll, J. J., and Solano, E.}
  \REVIEW{Phys. Rev. Lett.}{105}{2010}{263603}.

\bibitem{InnsbruckReview}
\Name{H\"affner, H., Roos, C. F., \and Blatt, R.}
  \REVIEW{Phys. Rep.}{469}{2008}{155}.

\bibitem{NISTreview03}
\Name{Leibfried, D., Blatt, R., Monroe, C., \and Wineland, D.}
  \REVIEW{Rev. Mod. Phys.}{75}{2003}{281}.

\bibitem{Gerritsma10}
\Name{Gerritsma, R., Kirchmair, G., Z\"ahringer, F., Solano, E., Blatt, R., and Roos, C. F.}
  \REVIEW{Nature}{463}{2010}{68}.

\bibitem{hwang2010}
\Name{Hwang, M. J. and Choi M. S.}
  \REVIEW{Phys. Rev. A}{82}{2010}{025802}.

\bibitem{Kyaw2014}
\Name{Kyaw, T. H., Felicetti, S., Romero, G., Solano, E., \and Kwek, L. C.}
  \REVIEW{Sci. Rep.}{5}{2015}{8621}.

\bibitem{messiah}
  \Name{Messiah, A.}
  \Book{Quantum Mechanics}
  \Publ{Dover Publications}
  \Year{1999}.

\bibitem{DAQS}
\Name{Lamata, L., Parra-Rodriguez, A., Sanz, M., \and Solano, E.}
  \REVIEW{Adv. Phys. X}{3}{2018}{1457981}.

\bibitem{Koch07}
\Name{Koch, J., Yu, T. M., Gambetta, J., Houck, A. A., Schuster, D. I., Majer, J., Blais, A., Devoret, M. H., Girvin, S. M., \and Schoelkopf, R. J.}
  \REVIEW{Phys. Rev. A}{76}{2007}{042319}.

\bibitem{Blais04}
\Name{Blais, A., Huang,  R.-S., Wallraff, A., Girvin, S. M. \and Schoelkopf, R. J.}
  \REVIEW{Phys. Rev. A}{69}{2004}{062320}.


\end{thebibliography}
\end{document}